\documentstyle[aps,epsfig,amsmath,twocolumn]{revtex}

\begin{document}

\title{\hfill TPJU-4/2001\\~~
\\Pion light cone wave function in the non-local NJL model}
\author{\textbf{Micha{\l} Prasza{\l}owicz and Andrzej Rostworowski}\\
\emph{M.Smoluchowski Institute of Physics,} \\\emph{Jagellonian University,} \\
\emph{Reymonta 4, 30-059 Krak\'{o}w, Poland.}}
\date{\today}
\maketitle
\begin{abstract}
We use the simple instanton motivated NJL-type model to calculate the leading
twist pion light cone wave function. The model consists in employing the
momentum dependent quark mass in the quark loop entering the definition of the
wave function. The result is analytical up to a solution of a certain
algebraic equation. Various properties including the $k_{T}$ dependence of the
pion wave function are discussed. The resulting $k_{T}$ integrated wave
function is not asymptotic and is in agreement with recent analysis of the
CLEO data.
\end{abstract}

\pacs{PACS:12.38.Lg, 13.60.Fz}

\section{Introduction} \label{intro}

Despite the fact that the exclusive hadron light cone wave functions were
theoretically introduced 20 years ago \cite{ChZh}\nocite{BrLep,FarJack}-
\cite{EfRad}, there is still relatively little experimental data which could
be confronted with theory. Recently the analysis of Ref.\cite{YakSch} based on
the latest CLEO measurements \cite{Cleo} put some limits on the expansion
coefficients of the pion wave function in terms of the Gegenbauer polynomials.
This analysis indicates that the pion wave function measured at $Q^{2}%
=1.5-9.2$ GeV$^{2}$ is neither asymptotic
\begin{equation}
\phi_{\pi}^{\mathrm{as}}(u)=6u(1-u), \label{as}%
\end{equation}
(with $u$ being the fraction of the pion momentum carried by the quark) nor of
the form proposed by Chernyak and Zhitnitsky in 1982 \cite{ChZh1,ChZhPRep}.
The first attempt to study the $k_{T}$ dependence of the pion wave function has
been undertaken in Refs.\cite{E791}.

In order to estimate the shape of the pion wave function at sub-asymptotic
$Q^{2}$, practically one theoretical approach has been used: QCD sum rules
\cite{SVZ} which gave a variety of contradictory results \cite{ChZhPRep,BrFil}%
. The extension of the local sum rules to the case of the non-zero quark
virtuality proposed in Refs.\cite{nonloc} has been recently successfully
applied to model the pion wave function \cite{BaMiSte}, which turned out to be
in agreement with the experimental analysis of Ref.\cite{YakSch}.

There have been also lattice calculations of the moments of the
pion wave function \cite{lattice1,lattice2} which, however, gave
unrealistically small results. Recent transverse lattice results
of Burkhardt et. al. \cite{Burk} and Dalley \cite{Dal} are in
contradiction. Dalley's pion wave function is much more flat than
the one of Ref.\cite{Burk}. It is, however, not very different
from our results for the constituent quark mass $M=325$ MeV.

Another approach has been recently put forward in Refs.\cite{PetPob,Bochum}
where a simple instanton based model was applied to calculate the
$u$-dependence of the pion and also photon wave function. The whole idea
consisted in calculating the quark loop which enters the expression for the
wave function, using the momentum dependent constituent quark mass. The
momentum dependence of the quark mass appears naturally in the instanton model
of the QCD vacuum \cite{DP}. It embodies the fact that the propagator of a
light quark gets modified by the vacuum through which the quark propagates.
The momentum dependent quark mass acts as a natural cut-off function for the
$k_{T}$ integration making this integration finite. Additionally, due to the
non-local $\pi$-quark coupling, it ensures vanishing of the pion wave function
at the end points (see Sect.\ref{san}). If the constituent quark mass is taken
as a constant and the sharp cut-off for the $k_T^2$ integration is used, the
pion wave function is simply a constant. For more sophisticated cut-offs one
can obtain more realistic shapes, as in Refs.\cite{TH}.

The instanton model of the QCD vacuum (for recent review see e.g.
\cite{DPrev}) allows to derive an effective non-local Nambu--Jona-Lasinio
(NJL) type quark-meson Lagrangian where mesons are not dynamical but composite
quark bilinears. By integrating out the quark fields one can, however,
generate the kinetic term for the mesons. Quark-meson (pion in our case)
interaction generated by the instantons is non-local and this non-locality is
parameterized by the momentum dependence of the constituent quark mass. The
constituent quark mass appears because of the chiral symmetry breaking induced
by the instantons. However, the instanton vacuum is most probably not able to
explain the confinement.

In most applications of the effective quark-meson theory described above an
approximation of the constant constituent quark mass was used. This approach
has an impressive success in describing baryon (which emerge as solitons)
properties (see \cite{DPrev} and references therein). Recently an attempt to
include momentum dependence in calculating solitons in the non-local NJL model
has been reported in Refs.\cite{BroRipGol}.

The instanton model of the QCD vacuum is naturally formulated in Euclidean
space-time, whereas the light cone wave function is an object defined in the
Minkowski space-time. One can in principle Wick rotate the underlying
integrals and perform the final integrals in the Euclidean space-time as has
been for example done in Refs.\cite{ET}-\nocite{DT}\cite{ADT} (for early
attempts to calculate the pion wave function in the instanton model see
Ref.\cite{Ryskin}). We will, however, perform all calculations directly
in the Minkowski space-time. The authors of Refs.\cite{PetPob,Bochum} used a
simple dipole type Ansatz for the momentum dependence of the constituent
quark mass in the Minkowski space-time which made the calculations feasible.
In Ref.\cite{PetPob} an approximation was used where the momentum dependence was taken into account only in the quark-pion vertices, whereas in
Ref.\cite{Bochum} the calculation has been done numerically. The results of
both calculations showed that the pion wave function at low momentum scale was
actually not very different from the asymptotic one of Eq.(\ref{as}).

In this paper we follow the approach of Refs.\cite{PetPob,Bochum}, however, we
show how to calculate the pion wave function \emph{analytically} (up to the
numerical solution of a certain algebraic equation). We obtain compact,
analytical results both for $u$ and $k_{T}^{2}$ dependent wave functions. We
use two parameter set of Ans\"{a}tze for the momentum dependent quark mass which
generalize the Ansatz of Refs.\cite{PetPob,Bochum}. In this way we can study
the sensitivity of various properties of the pion wave function to the
specific form of the cut-off function. This is quite important, since -- as
explained above -- we are not able to use the momentum dependent quark mass
calculated in the instanton model, but we model it in such a way that the
calculations can be performed in the Minkowski space-time.

A novelty of our approach is that we study the $k_{T}^{2}$ dependence of the
wave function. We calculate the first moments in $k_{T}^{2}$ which are related
to the mixed quark-gluon condensates \cite{Zhitkt}.

Our findings can be summarized as follows. The $d^{2}k_{T}$ integrated pion
wave function is not asymptotic, it exhibits a plateau for $u\sim1/2$ whose
height varies with the cut-off function and the constituent quark mass at zero
momentum. Comparing the coefficients of the expansion in terms of the
Gegenbauer polynomials, $a_{2k}$, with the analysis of the CLEO data
\cite{YakSch} we find that our wave function fits within 95\% confidence
level in the $a_{2}-a_{4}$ two dimensional parameter space, whereas the
asymptotic and the Chernyak-Zhitnitsky wave functions are ruled out. One
should note, however, that in the most cases the coefficient $a_{6}$ is in our
case not negligible, while it was explicitly put to zero in Ref.\cite{YakSch}.
In this respect we also differ from the model of Ref.\cite{BaMiSte}.

The $k_{T}^{2}$ dependence of the $u$ integrated wave function, for $k_{T}%
^{2}$ below $1$ GeV${}^{2}$, is insensitive to the considered forms of the
cut-off function. In this region, the wave function can be well fit by
$C(k_{T}^{2}+\Delta^{2})^{-\alpha}$ with $\alpha\sim8\div9$. For $k_{T}%
^{2}<0.2$ GeV${}^{2}$ the exponent fit can be used as well. The asymptotic
power like behavior depends strongly upon the form of the cut-off function.
However, the difference can be seen only for $k_{T}^{2}$ above few GeV${}^{2}%
$, whereas the region below $1$ GeV${}^{2}$ contributes 98\% to the norm of
the wave function. It indicates that the Fermilab experiment E791 \cite{E791}
where the typical quark (jet) $k_{T}$ is above $1$ GeV, is in fact sensitive
to the tiny portion of the pion wave function. The ratio $R=<k_{T}^{4}%
>/<k_{T}^{2}>^{2}$ is of the order of 3 indicating rather narrow $k_{T}^{2}$
distribution in contrast to Ref.\cite{Zhitkt} where $R\sim5-7$.

We find that, given simplicity of the model, both conceptual and technical,
our results are reasonable and, where the comparison with the data is
possible, surprisingly accurate. The method used in this paper can be
straightforwardly generalized to other wave functions like, for example,
photon wave function (which we briefly discuss in Sect.\ref{san}) or 2$\pi$
distribution amplitude.

The paper is organized as follows: in Section \ref{s2} we show how to
calculate the quark loop with the momentum dependent mass analytically. The
main problem here is, how to handle the complex poles of the quark propagator.
In Section \ref{s3} we discuss various properties of the pion wave function,
like end point behavior and $k_{T}$ asymptotics. We also present numerical
results and compare with other models. Finally in Section \ref{sumc} we
present our conclusions.

\section{Pion wave function with momentum dependent quark mass} \label{s2}

Let us consider two quarks moving along the light cone direction $\tilde
{n}=(1,0,0,1)$ parallel to the total momentum $P$ and separated by the light
cone distance $z=2\tau$ along the direction $n=(1,0,0,-1)$. The twist 2 light
cone pion wave function is then defined as
\begin{align}
\phi_{\pi}(u)  &  =\frac{1}{i\sqrt{2}F_{\pi}}\int\limits_{-\infty}^{\infty
}\frac{d\tau}{\pi}e^{-i\tau(2u-1)(nP)}\nonumber\\
&  \left\langle 0\right|  \bar{\psi}(n\tau)\rlap{/}n\gamma_{5}\psi\left(
-n\tau\right)  \left|  \pi^{+}(P)\right\rangle \label{Fipidef}%
\end{align}
where variable $u$, as will be shortly seen, has a meaning of a fraction of a
quark momentum $k^{+}$ with respect to $P^{+}$. Here $F_{\pi}=93$ MeV. In this
kinematical frame any four vector $v$ can be decomposed as:
\begin{equation}
v^{\mu}=\frac{v^{+}}{2}\tilde{n}^{\mu}+\frac{v^{-}}{2}n^{\mu}+v_{T}^{\mu}
\label{LC}%
\end{equation}
with\quad$v^{+}=n\cdot v,\quad v^{-}=\tilde{n}\cdot v$ and the scalar product
of two four vectors reads:
\begin{equation}
v\cdot w=\frac{1}{2}v^{+}w^{-}+\frac{1}{2}v^{-}w^{+}-\vec{v}_{T}\cdot\vec
{w}_{T}.
\end{equation}
In Eq.(\ref{Fipidef}) the path ordered exponential of the gluon field,
required by the gauge invariance, has been omitted since we shall be working
in the effective quark model where the gluon fields have been integrated out.

The instanton model of the QCD vacuum predicts that quarks interact
\emph{non-locally} with an external meson field $U$ \cite{DP,DPrev}%
\begin{equation}
S_{I}=M\int\frac{d^{4}kd^{4}l}{(2\pi)^{8}}\bar{\psi}(k)F(k)U^{\gamma_{5}%
}(k-l)F(l)\psi(l)\, \label{SI}%
\end{equation}
and $U^{\gamma_{5}}(x)$ can be expanded in terms of the pion fields:
\begin{equation}
U^{\gamma_{5}}=1+\frac{i}{F_{\pi}}\gamma^{5}\tau^{A}\pi^{A}-\frac{1}{2F_{\pi
}^{2}}\pi^{A}\pi^{A}+\ldots\label{U}%
\end{equation}
$M$ is a constituent quark mass of the order of $350$ MeV and $F(k)$ is a
momentum dependent function such that $F(0)=1$ and $F(k^{2}\rightarrow
\infty)\rightarrow0$.
%
%
\begin{figure}[h]
\centerline{\epsfig{file=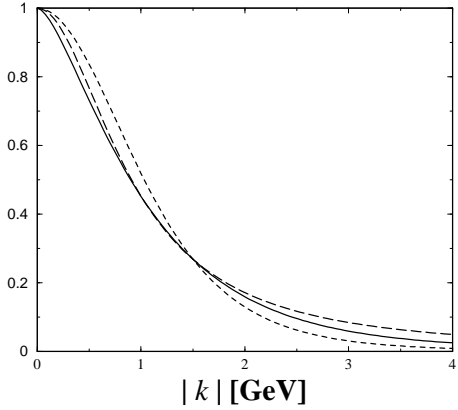,width=6cm}} \caption{{\small $F(k)$ for
Euclidean momentum $k^{2}<0$, for $n=1$ (long dashed), 3 (dashed)
and for the instanton model (solid).}}%
\label{cutoff}%
\end{figure}

Function $F(k)$ has been calculated in the instanton model of the QCD vacuum
\begin{equation}
F_{inst.}(k)=2z\left[  I_{0}(z)K_{1}(z)-I_{1}(z)K_{0}(z)\right]
-2I_{1}(z)K_{1}(z) \label{Fkinst}%
\end{equation}
where $z=k\rho/2$. Here $1/\rho\sim600$ Mev is a typical inverse instanton
size. One has to remember that this function is defined in Euclidean space as
a Fourier transform of the instanton zero mode. Here, following
Refs.\cite{PetPob,Bochum} we wish to calculate $\phi_{\pi}$ directly in the
Minkowski space. To this end we shall choose a simple pole formula
\begin{equation}
M(k)=MF^{2}(k)\qquad\text{with}\qquad F(k)=\left(  \frac{-\Lambda^{2}}%
{k^{2}-\Lambda^{2}+i\epsilon}\right)  ^{n} \label{Fkdef}%
\end{equation}
which reproduces reasonably well (\ref{Fkinst}) for $k^{2}<0$. This is shown
in Fig.\ref{cutoff} where we have chosen $\Lambda$ in such a way that
(\ref{Fkdef}) crosses with (\ref{Fkinst}) for $k=1.5$ GeV. Unfortunately large
$k$ asymptotics $F_{inst}(k)\sim k^{-3}$ cannot be reproduced by
Eq.(\ref{Fkdef}) for integer $n$. Choosing half-integer $n$, however, would
introduce cuts in the complex $k$ plane, making the numerical evaluation of
the pion wave function unnecessary tedious. The cutoff parameter $\Lambda$ has
to be chosen in such a way that the pion wave function satisfies the
normalization condition
\begin{equation}
\int\limits_{0}^{1}du\,\phi_{\pi}(u)=1. \label{Phinorm}%
\end{equation}

Matrix element (\ref{Fipidef}) reduces to a simple fermion loop
\begin{align}
\phi_{\pi}(u) &  =-\frac{iN_{c}}{F_{\pi}^{2}P^{+}}\int\frac{d^{4}k}{\left(
2\pi\right)  ^{4}}\,\delta\left(  u-\frac{k_{+}}{P_{+}}\right)  \nonumber\\
&  {\mathrm Tr}\left[  \rlap{/}{n}\,\gamma_{5}\frac{i\sqrt{M(k)}}{\rlap
{/}k-M(k)}\gamma_{5}\frac{i\sqrt{M(k-P)}}{\left(  \rlap{/}k-\rlap{/}%
{P}\right)  -M(k)}\right]  .\label{PhiTr}%
\end{align}
Here it is explicitly seen that $uP_{+}=k_{+}$. In the light-cone
representation (\ref{LC}) $d^{4}k=\frac{1}{2}dk^{+}dk^{-}d^{2}\vec{k}^{\bot}$
and after performing the trace we get
\[
\phi_{\pi}(u)=-i\frac{2N_{c}M^{2}P_{+}}{F_{\pi}^{2}}\int\frac{d^{2}k_{T}%
}{\left(  2\pi\right)  ^{2}}\int\frac{dk_{-}}{\left(  2\pi\right)  ^{2}%
}\,\,F(k)F(k-P)
\]%
\begin{equation}
\frac{(1-u)F^{2}(k)+u\,F^{2}(k-P)}{[k^{2}-M^{2}F^{4}(k)+i\epsilon][\left(
k-P\right)  ^{2}-M^{2}F^{4}(k-P)+i\epsilon]}\label{Fpi1}%
\end{equation}

In order to calculate the integral over $dk_{-}$ \emph{exactly} i.e. with full
$k$ dependence in the denominators we have to find zeros of the two
propagators in Eq.(\ref{Fpi1}). To this end it is convenient to introduce the
scaled variables
\begin{align}
p_{+} &  =\frac{P_{+}}{\Lambda},\quad\kappa_{-}=\frac{k_{-}}{\Lambda}%
,\quad\quad\vec{\kappa}_{T}=\frac{\vec{k}_{T}}{\Lambda},\quad\nonumber\\
\eta &  =p_{+}\kappa_{-},\quad t=(\vec{\kappa}_{T})^{2},\quad r^{2}%
=\frac{M^{2}}{\Lambda^{2}}.\label{sc}%
\end{align}
In these variables
\begin{align}
\phi_{\pi}(u)  & =-i\frac{N_{c}M^{2}}{(2\pi)^{3}F_{\pi}^{2}}\int_{0}^{+\infty
}dt\int_{-\infty}^{+\infty}d\eta\,\nonumber\\
& \frac{(1-u)\,\zeta_{1}^{n}\,\zeta_{2}^{3n}+u\,\,\zeta_{1}^{3n}\,\zeta
_{2}^{n}}{G\left(  \zeta_{1}\right)  \,G\left(  \zeta_{2}\right)
}\label{piwfz}%
\end{align}
where
\begin{align}
\zeta_{1} &  =u\left(  \eta-\frac{t+1}{u}+i\epsilon\,\mbox{sign}(u)\right)
,\quad\nonumber\\
\zeta_{2} &  =-(1-u)\left(  \eta+\frac{t+1}{(1-u)}-i\epsilon\,\mbox
{sign}(1-u)\right)  \label{zzbar}%
\end{align}
and
\begin{equation}
G(z)=z^{4n+1}+z^{4n}-r^{2}=\prod\limits_{i=1}^{4n+1}\left(  z-z_{i}\right)
,\label{Gz}%
\end{equation}
with $z_{i}$ being $(4n+1)$ roots of the equation
\begin{equation}
G(z)=0.\label{Geq0}%
\end{equation}
We aim at evaluating the $d\eta$ integral by contour integration. The
$2(4n+1)$ simple poles of the integrand (\ref{piwfz}) lie at
\begin{align}
\eta_{i}^{(1)} &  =\frac{t+1+z_{i}}{u}-i\epsilon\,\mbox{sign}(u),\nonumber\\
\eta_{i}^{(2)} &  =-\frac{t+1+z_{i}}{1-u}+i\epsilon\,\mbox{sign}(1-u).
\end{align}

\begin{figure}[h]
\begin{center}
\epsfig{file=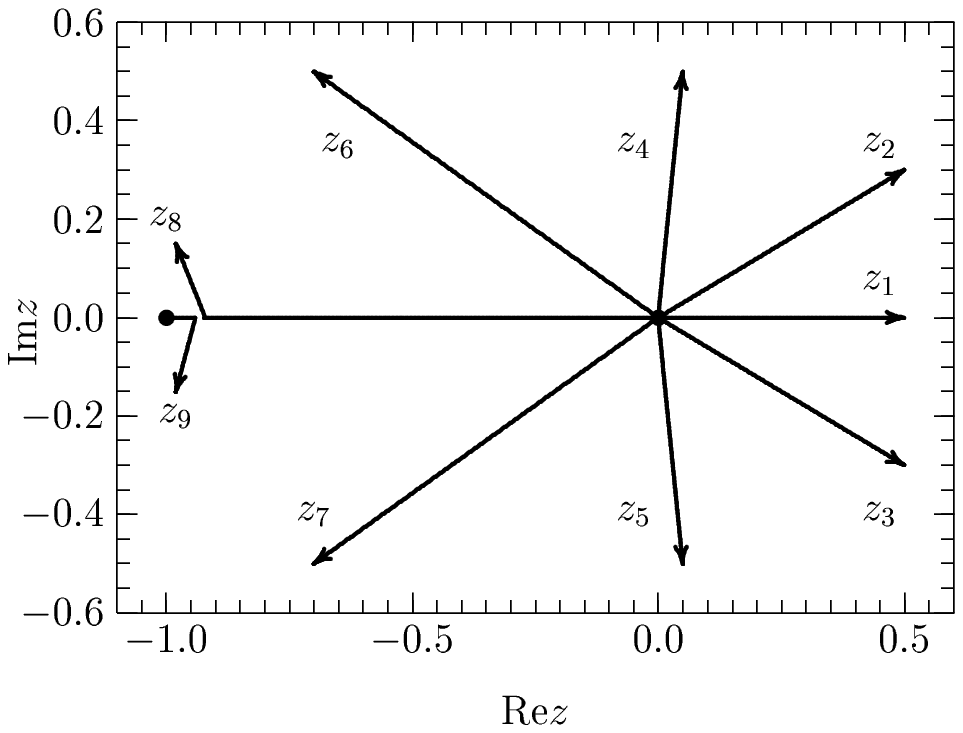,width=7cm}
\end{center}
\caption{{\small Flow of the zeros of $G(z)$with increasing $r$ for $n=2$. For
$r=0$ $(\Lambda=\infty)$ there are $4n$ degenerate solutions $z_{1}%
,...,z_{4n}$ and one non degenerate solution $z_{4n+1}=-1$. With increasing $r$
($\Lambda$ decreasing from $\infty$ to the physical value) $z_{4n+1}$ solution
drifts rightwards toward $0$, and $4n$ zero solutions split into 2 real ones,
$z_{4n}$ drifting towards $z_{4n+1}$ and $z_{1}$ towards $+\infty$, and $4n-2$
complex (pairwise conjugated) solutions $z_{2i},z_{2i+1}$, $i=1,...,2n-1$
which flow into the upper and lower parts of the complex $z$ plane. When $r$
reaches the value of the local maximum $r_{\max}$ of $G(z)$ for $-1<z<0$ the
two real zeros $z_{4n}$ and $z_{4n+1}$ meet and ''scatter'' drifting further
into the upper ($z_{4n}$) and lower ($z_{4n+1}$) half of the complex $z$
plane. }}%
\label{fig:zzeros}%
\end{figure}

\begin{figure}[hh]
\begin{center}
\epsfig{file=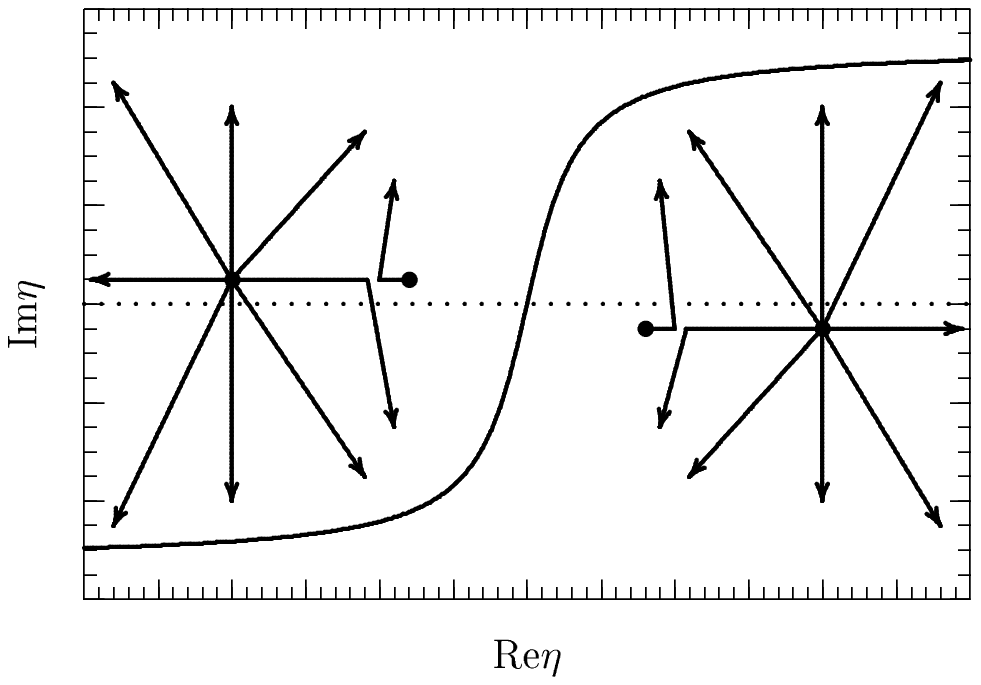,width=7cm}
\end{center}
\caption{{\small Schematic flow of the $\eta$ poles with increasing $r$ for
$n=2$ together with the integration contour in the complex $\eta$ plane. For
$r=0$ the poles are shown as black dots and the integration contour is
represented by the dotted line.}}%
\label{ktpoles}%
\end{figure}

Let's assume for definiteness that $0\leq u\leq1$. For
$r^{2}=0$ ($\Lambda^{2}=\infty$) the roots of Eq. (\ref{Geq0}) are
$z_{1}=...=z_{4n}=0$, $z_{4n+1}=-1$ so the poles $\eta^{(1)}$ lie below and
the poles $\eta^{(2)}$ lie above the Re$\eta$ axis. When $r^{2}$ increases
($\Lambda^{2}$ decreases from $\infty$ to its physical value) the roots
$z_{i}$ drift in the complex z-plane as shown in Fig. \ref{fig:zzeros}.
Therefore some of the poles $\eta_{i}^{(1)}$ can move above and some of the
poles $\eta_{i}^{(2)}$ can move below the Re$\eta$ axis. The \textit{crucial}
point is that if we kept the integration path going along the Re$\eta$ axis we
would get an unphysical result: $\phi_{\pi}(u)$ would not vanish outside the
interval $0\leq u\leq1$. We assume that physically correct prescription for
performing the $d\eta$ integral is to deform the integration path as shown in
Fig. \ref{ktpoles}.

Closing the contour in the upper complex $\eta$ half-plane
we should enclose all the poles $\eta_{i}^{(2)}$ and none of the poles
$\eta_{i}^{(1)}$. This prescription ensures that $\phi_{\pi}(u)$ is real and
vanishes outside the interval $0\leq u\leq1$. Now, the $d\eta$ integral
yields\newline
\begin{align}
\phi_{\pi}(u) &  =\frac{N_{c}M^{2}}{(2\pi)^{2}F_{\pi}^{2}}(-1+u)^{n}\int
_{0}^{+\infty}dt\,\sum_{i=1}^{4n+1}z_{i}^{n}f_{i}\\
&  \frac{u(t+1+uz_{i})^{3n}+(1-u)^{2n+1}z_{i}^{2n}(t+1+uz_{i})^{n}}%
{\prod_{k=1}^{4n+1}[t+1+uz_{i}+(1-u)z_{k}]},\nonumber
\end{align}
where factors $f_{i}$ are defined as
\begin{equation}
f_{i}=\prod\limits_{{}_{\scriptstyle k\neq i}^{\scriptstyle k=1}}^{4n+1}%
\frac{1}{z_{k}-z_{i}}.\label{fidef}%
\end{equation}
With the help of the following identity
\begin{align}
\label{id1}
&  \frac{u(t+1+uz_{i})^{3n}+(1-u)^{2n+1}z_{i}^{2n}(t+1+uz_{i})^{n}}%
{\prod_{k=1}^{4n+1}[t+1+uz_{i}+(1-u)z_{k}]}\nonumber\\
&  =\frac{1}{(-1+u)^{n}}\sum_{k=1}^{4n+1}z_{k}^{n}f_{k}\frac{uz_{k}%
^{2n}+(1-u)z_{i}^{2n}}{t+1+uz_{i}+(1-u)z_{k}}%
\end{align}
the result for the $k_{T}^{2}$-dependent pion wave function $\Phi(u,k_{T}%
^{2})$ can be cast in the following compact form:
\begin{align}
\Phi_{\pi}(u,k_{T}^{2})  & =\frac{1}{\Lambda^{2}}\frac{N_{c}M^{2}}{(2\pi
)^{2}F_{\pi}^{2}}\nonumber\\
& \sum\limits_{i,k=1}^{4n+1}f_{i}f_{k}\,\frac{z_{i}^{n}z_{k}^{3n}u+z_{i}%
^{3n}z_{k}^{n}(1-u)}{\frac{k_{T}^{2}}{\Lambda^{2}}+1+z_{i}u+z_{k}%
(1-u)}.\label{Phiktu}%
\end{align}
Factors $f_{i}$ obey the following properties which are crucial for the
finiteness of the $d(k_{T}^{2})$ integration and for the end point behavior of
the pion (and also photon) wave function
\begin{equation}
\sum\limits_{i=1}^{4n+1}z_{i}^{m}f_{i}=\left\{
\begin{array}
[c]{ccc}%
0 & \text{for} & m<4n,\\
&  & \\
1 & \text{for} & m=4n.
\end{array}
\right.  \label{fiprop}%
\end{equation}
It is interesting to note that properties (\ref{fiprop}) hold for any set
of $N=4n+1$ numbers, irrespectively of the fact that they are solutions of
certain equation of degree $N$.

Due to Eq.(\ref{fiprop}) $\Phi_{\pi}(u,k_{T}^{2})$ vanishes for large
$k_{T}^{2}$ and the $d(k_{T}^{2})$ integration is finite so that
\begin{align}
\phi_{\pi}(u) &  =\int\limits_{0}^{\infty}d(k_{T}^{2})\,\Phi_{\pi}(u,k_{T}%
^{2}) \label{Phiu} 
\nonumber\\
& = -\frac{N_{c}M^{2}}{(2\pi)^{2}F_{\pi}^{2}}\sum\limits_{i,k}f_{i}%
f_{k} (z_{i}^{n}z_{k}^{3n}u+z_{i}^{3n}z_{k}^{n}(1-u))
\nonumber\\
& \times \ln\left(1+z_{i}u+z_{k}(1-u)\right).
\end{align}

Eqs.(\ref{Phiktu},\ref{Phiu}) are our final results for the pion light cone
wave function for the momentum dependent quark mass (\ref{Fkdef}). These
analytical formulae depend on the numerical solutions $z_{i}$ of
Eq.(\ref{Geq0}). Cutoff parameter $\Lambda$ can be found by imposing the
normalization condition (\ref{Phinorm}).

\section{Properties of the pion wave function} \label{s3}

\subsection{Analytical properties} \label{san}

Function $\phi_{\pi}(u)=\phi_{\pi}(1-u)$ which is a trivial property of the
symmetry of exchanging two sums over $i$ and over $k$ in Eq.(\ref{Phiu}). At
the end points $\phi_{\pi}(u)$ vanishes. Indeed
\begin{equation}
\phi_{\pi}(0)=-\frac{N_{c}M^{2}}{(2\pi)^{2}F_{\pi}^{2}}\sum\limits_{i}%
f_{i}z_{i}^{3n}\sum\limits_{k}f_{k}z_{k}^{n} \ln\left(1+z_{k} \right)=0
\end{equation}
as a consequence of Eq.(\ref{fiprop}). In fact for small $u$ $\phi_{\pi
}(u)\sim u^{n}$. The reason for this kind of behavior is the presence of the
factor $\sqrt{M(k)M(k-P)}$ in Eq.(\ref{PhiTr}). Indeed, one can calculate in
the same way the transverse photon wave function $\phi_{\gamma}^{\bot}(u)$
which, since the $\gamma$-quark coupling is local, does not contain the
dumping factor $\sqrt{M(k)M(k-P)}$. The result reads
\begin{align}
\phi_{\gamma}^{\bot}(u) & =-\frac{N_{c}M}{(2\pi)^{2}f_{\gamma}^{\bot}}%
\sum\limits_{i,k}f_{i}f_{k} (z_{i}^{2n}z_{k}^{4n}u+z_{i}^{4n}z_{k}%
^{2n}(1-u))
\nonumber\\
& \times \ln\left(  1+z_{i}u+z_{k}(1-u)\right)%
\end{align}
and the wave function $\phi_{\gamma}^{\bot}(0)=\phi_{\gamma}^{\bot}(1)\neq0$,
due to the second property (\ref{fiprop}). So despite of the nonperturbative
effects embodied in the momentum dependent quark mass (\ref{Fkdef}) the
point like photon-quark coupling makes the photon wave function non vanishing at
the end points.

The advantage of the present calculation is that we can calculate not only the
$u$ dependence, but also the $k_{T}^{2}$ dependence of the pion wave function.
Integrating $\Phi_{\pi}(u,k_{T}^{2})$ over $u$ we get
\begin{equation}
\tilde{\phi}_{\pi}(k_{T}^{2})=\int\limits_{0}^{1}du\,\Phi_{\pi}(u,k_{T}^{2}).
\label{Phikt}%
\end{equation}
\begin{figure}[h]
\begin{center}
\epsfig{file=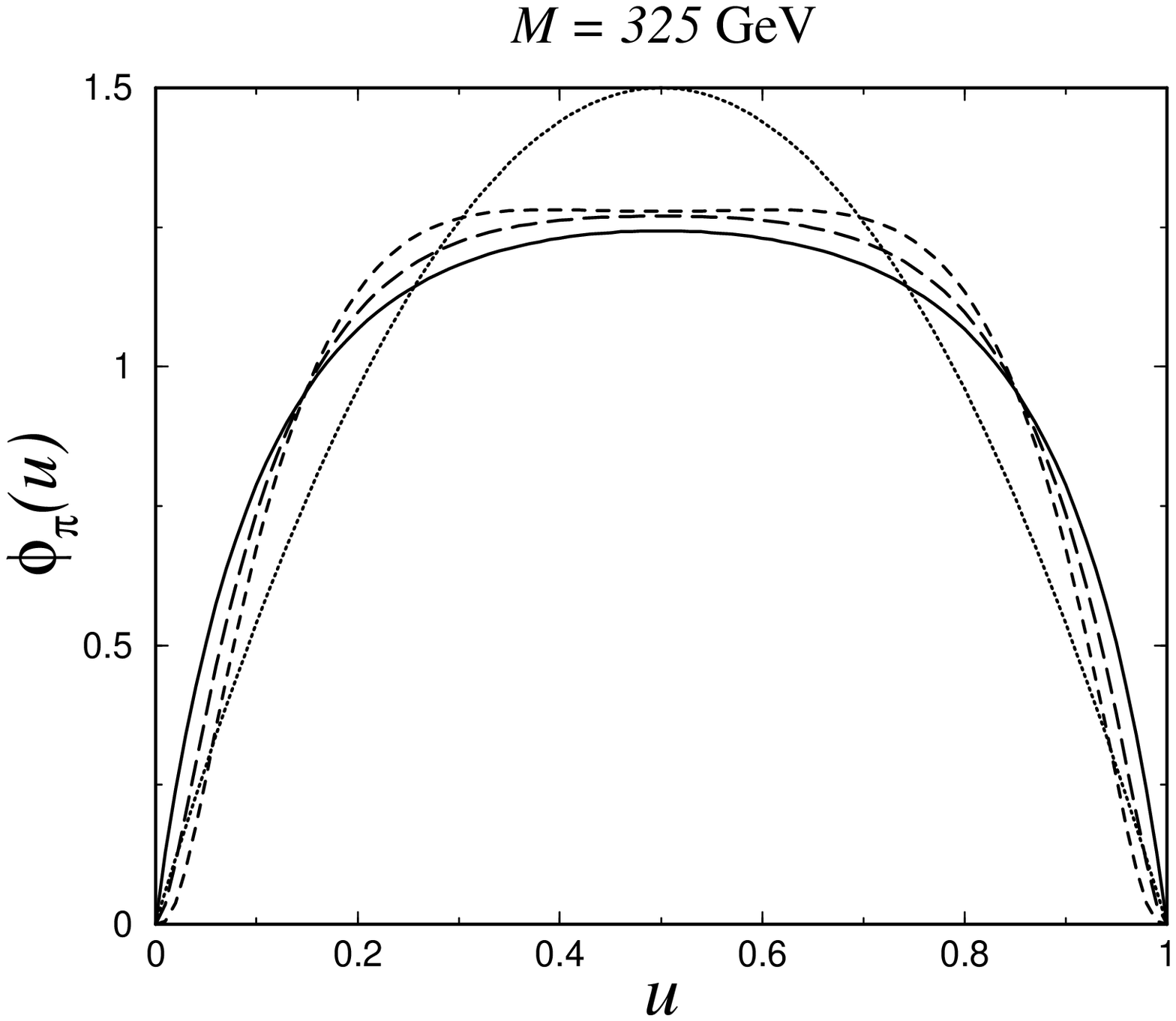,width=6cm} \\
\epsfig{file=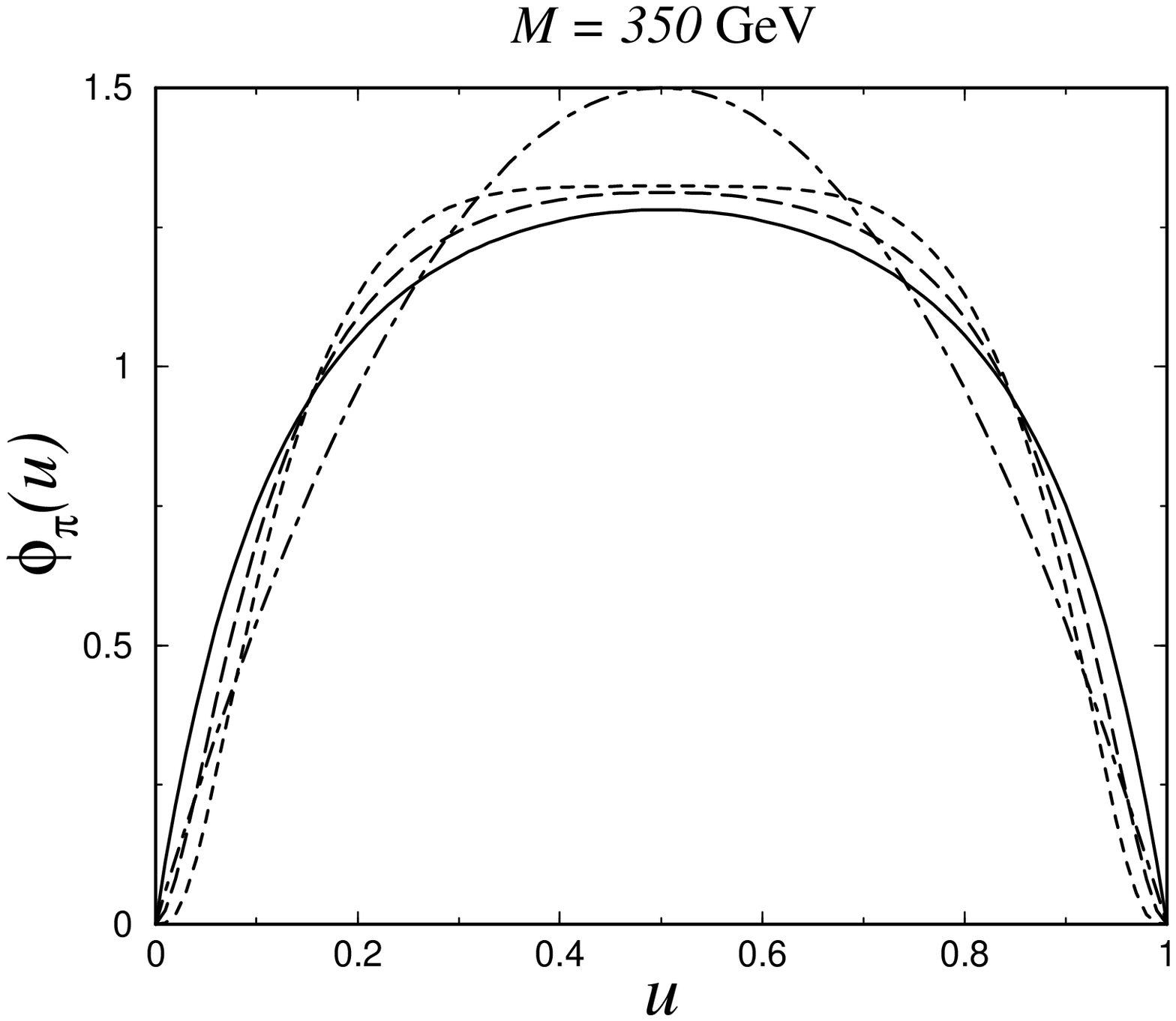,width=6cm} \\
\epsfig{file=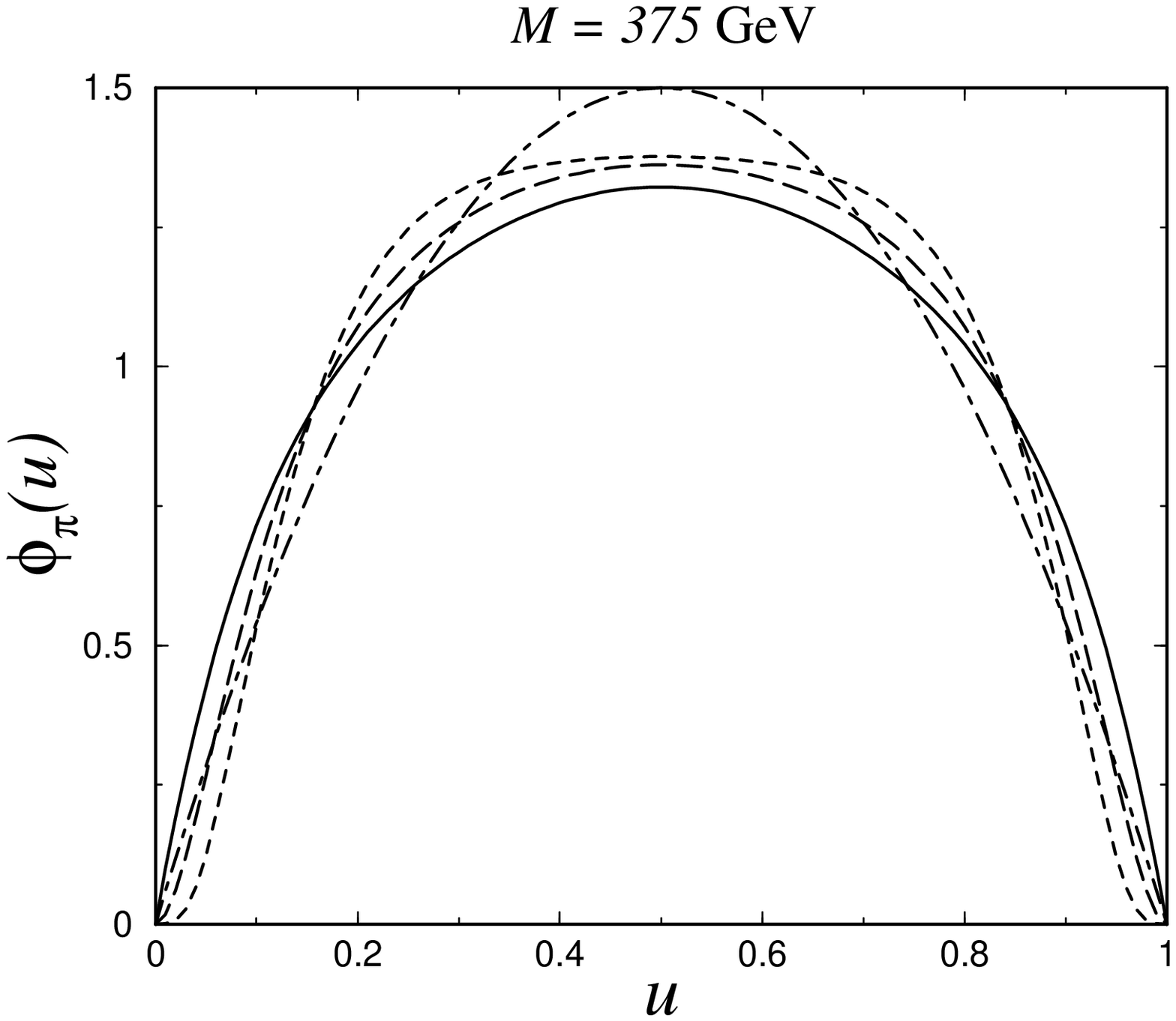,width=6cm} \\
\epsfig{file=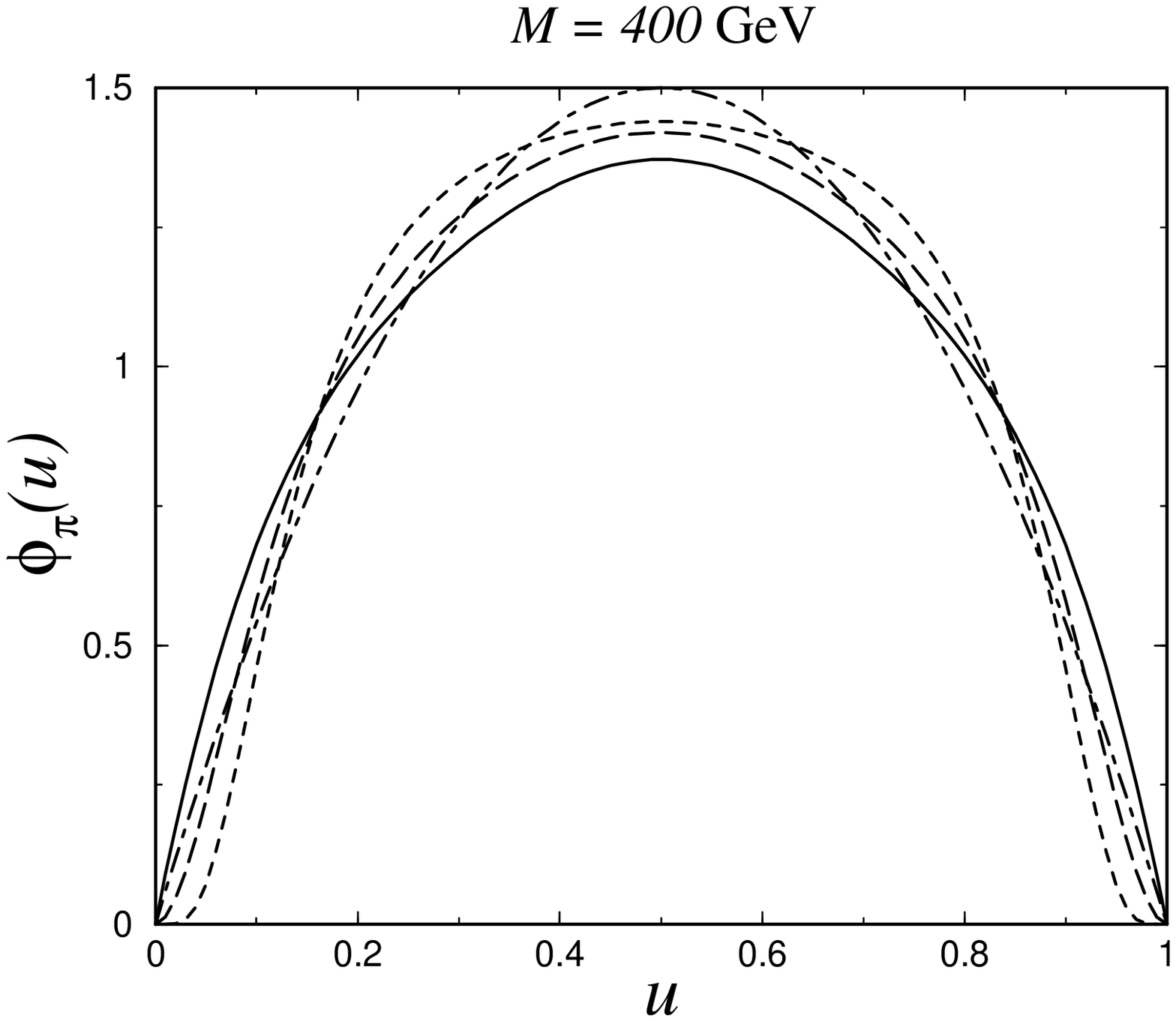,width=6cm}
\end{center}
\caption{Pion wave function for $M=325$, 350, 375 and 400~MeV, for
$n=1$ (solid), 2 (long dashed) and 5 (dashed) together with the
asymptotic wave
function (dot-dashed).}%
\label{allpwf}%
\end{figure}
By expanding $\tilde{\phi}(k_{T}^{2})$, Eq.(\ref{Phikt}), for large $k_{T}%
^{2}$ we find
\begin{equation}
\tilde{\phi}_{\pi}(k_{T}^{2})\sim\left(  \frac{\Lambda^{2}}{k_{T}^{2}}\right)
^{4n+1}. \label{Phiktas}%
\end{equation}
This kind of behavior differs from the commonly assumed exponential
form\footnote{The exponential fit, however, is quite satisfactory for
$k_{T}^{2}<0.2$ GeV$^{2}$.} of $\tilde{\phi}_{\pi}(k_{T}^{2})$. Let us,
however, remark that the asymptotic behavior (\ref{Phiktas}) which switches
on for $k_{T}^{2}>\Lambda^{2}$ concerns only a negligible tail of the whole
wave function. Moreover, the power in Eq.(\ref{Phiktas}) is large enough, even
for $n=1$, to make the predictions of the first $k_{T}^{2}$ moments reliable.
We shall come back to this point in the next Section.

\subsection{Numerical results \label{snum}}

In Fig.\ref{allpwf} we plot $\phi_{\pi}(u)$ for $M=325,350,375,400$ MeV and
$n=1,2$ and $5$. In fact for $n>5$ $\phi_{\pi}(u)$ does not change much
anymore. It is clearly seen from these figures that in all cases the NJL model
with the non-local regulator (\ref{Fkdef}) gives $\phi_{\pi}(u)$ which is
different from the asymptotic one. This difference is more pronounced for
smaller masses $M$ and larger $n$'s. For $M=325$ and $350$ Mev and $n=5$ a
shallow minimum can be seen at $u=0.5$.

\begin{figure}[h]
\begin{center}
\epsfig{file=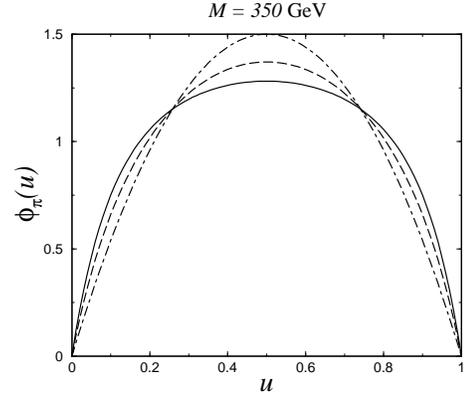,width=6cm}
\end{center}
\caption{Pion wave function for $M=350$~MeV, for $n=1$ (solid), compared with
the approximate result of Ref.[17] (dashed) and the asymptotic wave function
(dot-dashed).}%
\label{fig:comp}%
\end{figure}

It is instructive to compare our result with the one of
Refs.\cite{PetPob,Bochum} where the momentum dependence of the quark mass has
been taken into account only in the numerators of Eq.(\ref{Fpi1}). This is
done in Fig.\ref{fig:comp} for $M=350$ MeV and $n=1$. It is clearly seen that
the momentum dependence of the denominators flattens the pion wave function
making it ''less asymptotic''.
\begin{table}[h]
\caption{ Gegenbauer coefficients}%
\label{tb:gegen}
\begin{center}%
\begin{tabular}
[c]{cccrrrr}%
$M$ & n & $\Lambda$ & $a_{2}$ & $a_{4}$ & $a_{6}$ & $a_{8}$\\
MeV &  & MeV &  &  &  & \\\hline
& 1 & 1249 & 0.1363 & 0.0215 & 0.0037 & 0.0007\\
325 & 2 & 1862 & 0.0935 & -0.0183 & -0.0155 & -0.0077\\
& 5 & 3033 & 0.0606 & -0.0493 & -0.0217 & -0.0060\\\hline
& 1 & 1156 & 0.1144 & 0.0151 & 0.0015 & 0.0001\\
350 & 2 & 1727 & 0.0659 & -0.0271 & -0.0166 & -0.0064\\
& 5 & 2819 & 0.0284 & -0.0600 & -0.0202 & -0.0020\\\hline
& 1 & 1081 & 0.0931 & 0.0113 & -0.0003 & -0.0002\\
375 & 2 & 1621 & 0.0387 & -0.0325 & -0.0166 & -0.0047\\
& 5 & 2649 & -0.0035 & -0.0668 & -0.0173 & 0.0025\\\hline
& 1 & 1020 & 0.0723 & 0.0099 & -0.0020 & -0.0001\\
400 & 2 & 1534 & 0.0118 & -0.0347 & -0.0164 & -0.0027\\
& 5 & 2512 & -0.0351 & -0.0695 & -0.0138 & 0.0072
\end{tabular}
\end{center}
\end{table}

The comparison of different wave functions and especially their content of the
non-asymptotic terms is readily done by expanding $\phi_{\pi}(u)$ in terms of
the Gegenbauer polynomials:
\begin{align}
\phi_{\pi}(u)  & =\phi_{\pi}^{\text{as}}(u)\left[  1+a_{2}C_{2}^{3/2}%
(2u-1)+a_{4}C_{4}^{3/2}(2u-1)\right.  \nonumber\\
& \left.  +a_{6}C_{6}^{3/2}(2u-1)+\ldots\right]  \label{gegen}%
\end{align}
where the coefficients $a_{2k}$ depend on the scale $\mu^{2}$. In model
calculations it is, however, hard to define precisely what value should be
taken for $\mu$. Clearly $\mu$ is of the order of the cutoff $\Lambda$. In
Table \ref{tb:gegen} we show the cutoff $\Lambda$ and $a_{2},\ldots,a_{8}$ for
$M=325,350,375,400$ MeV and $n=1,2$ and $5$. It is clearly seen that in most
cases $a_{6}$ is still comparable with $a_{4}$. Only $a_{8}$ is always
substantially smaller than $a_{2},a_{4}$ or $a_{6}$. Interestingly for $M=325$
MeV and $n=1$ we get $a_{2}$ very similar to the transverse lattice result of
Dalley \cite{Dal} who gets $a_{2}^{latt.}=0.133$.

\begin{figure}[h]
\begin{center}
\epsfig{file=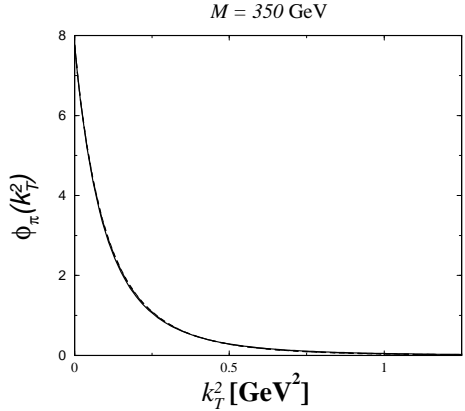,width=6cm}
\end{center}
\caption{$\tilde{\phi}_{\pi}(k_{T}^{2})$ for $M=350$ and $n=1,2$ and $5$.
These curves are indistinguishable within the scale of the plot.}%
\label{fig:kt}%
\end{figure}

As already stated in the previous sections the non-local NJL model gives the
pion wave function as a function of two variables $u$ and $k_{T}^{2}$. In
Fig.\ref{fig:kt} we plot $\tilde{\phi}_{\pi}(k_{T}^{2})$ defined in
Eq.(\ref{Phikt}) for $M=350$~Mev and $n=1,2$ and $5$. These functions are
indistinguishable within the accuracy of the plot. $\tilde{\phi}_{\pi}%
(k_{T}^{2})$ can be well approximated by the function $C(k_{T}^{2}+\Delta
^{2})^{-\alpha}$, with $\alpha=9,$ $\Delta=1$ GeV$^{2}$ and $C=8$
GeV$^{16}$. The $n$ dependence is visible only for $k_{T}^{2}$ above a few
GeV${}^{2}$ and is not taken into account in this fit.

In order to illustrate the (un)importance of the asymptotic power like
behavior, let us define
\begin{equation}
{\mathcal N}(\lambda)=\int\limits_{0}^{\lambda^{2}} d(k_{T}^{2})\,
\tilde{\phi}_{\pi}(k_{T}^{2}).
\end{equation}
Obviously ${\mathcal N}(\infty)=1$. However, $\mathcal{N}(\lambda)$ saturates
relatively fast
\begin{equation}
{\mathcal N}(1\;\text{GeV})\sim0.98
\end{equation}
for all masses and powers of $n$. In fact the higher $n$ the larger value of
${\mathcal N}(1\;$GeV$)$. This observation indicates that the bulk of the wave
function comes from the region where there is little, if any, dependence on
the cut-off parameter $n$. Outside this region the $k_{T}^{2}$-dependent wave
function starts to depend strongly on the form of the cutoff function
(\ref{Fkdef}). Unfortunately the $k_{T}$ range in the recently measured
diffractive dijet production in the pion nucleus scattering \cite{E791} is
precisely in this asymptotic regime. Therefore we have to conclude that the
information provided by this piece of data cannot be used to test our model,
unless the $k_{T}$ range is not shifted towards the MeV rather than GeV
region.
\begin{figure}[h]
\begin{center}
\epsfig{file=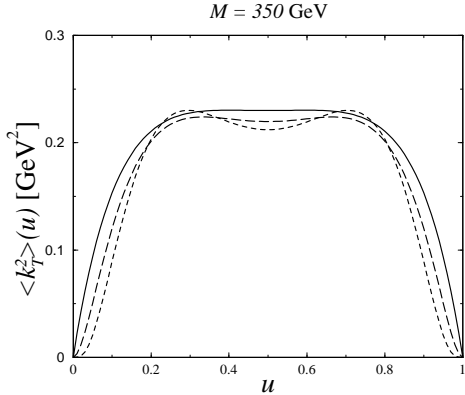,width=6cm}
\end{center}
\caption{$<k_{T}^{2}>$ in pion wave function for $M=350$~MeV, and $n=1$
(solid), 2 (long dashed) and 5 (dashed).}%
\label{fig:kt2av}%
\end{figure}

In order to see how the transverse momentum is distributed among quarks with
different longitudinal momentum it is instructive to plot $<k_{T}^{2}>$ as a
function of $u$. This is depicted in Fig.\ref{fig:kt2av}. One can see that
$<k_{T}^{2}>$ saturates relatively fast and remains almost constant for
$0.2<u<0.8$. The saturation value decreases with an increasing $M$. In Table
\ref{tb:ktav} average $<k_{T}^{2}>$ and $<k_{T}^{4}>$ are displayed together
with the ratio $R=<k_{T}^{2}>^{2}/<k_{T}^{4}>$. It was argued in
Ref.\cite{Zhitkt} that the QCD sum rules predict $R\sim4-5$. In our case $R$
is almost two times smaller, however the absolute value of $<k_{T}^{2}>$ is
slightly larger than the one of Ref.\cite{Zhitkt}. Let us remind that the
$k_{T}^{2}$ moments are related to the mixed quark-gluon condensates which are
poorly known in QCD.
\begin{figure}[h]
\begin{center}
\epsfig{file=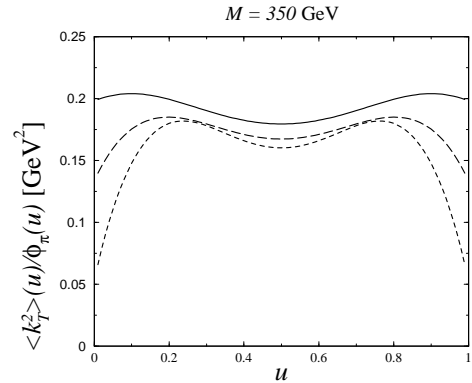,width=6cm}
\end{center}
\caption{Ratio $<k_{T}^{2}>(u)/\phi_{\pi}(u)$ for $M=350$~MeV, and $n=1$
(solid), 2 (long dashed) and 5 (dashed). If there was factorization, this
ratio should not depend upon $u$.}%
\label{fig:kt2fact}%
\end{figure}

We have checked the hypothesis that $\Phi(u, k_{T}^{2})$ factorizes:
\begin{equation}
\Phi_{\pi}(u, k_{T}^{2}) = \phi_{\pi}(u) \tilde\phi_{\pi}(k_{T}^{2}).
\end{equation}
If this was the case we should get $<k_{T}^{2}>(u)/\phi_{\pi}(u) =
\mbox{const}= <k_{T}^{2}>$. The actual plot of $<k_{T}^{2}>(u)/\phi_{\pi}(u)$
dependence is presented in Fig.\ref{fig:kt2fact}.

\subsection{Comparison with the data \label{sdata}}

\begin{figure}[h]
\begin{center}
\epsfig{file=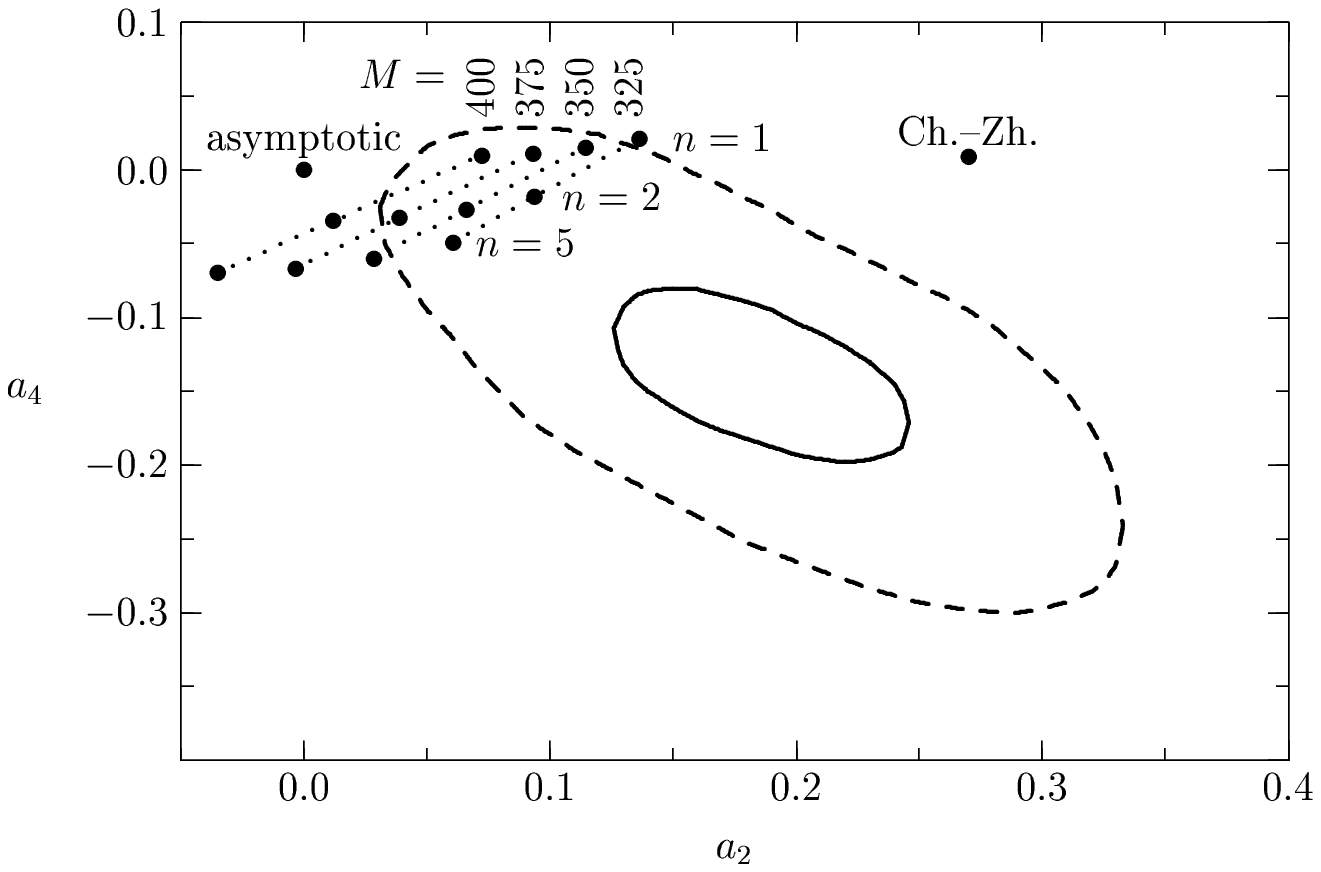,width=8.5cm}
\end{center}
\caption{The parameter space ($a_{2},a_{4}$) of Ref.[5]. Black dots represent
different model predictions, solid contour corresponds to 68\% confidence
level, whereas the dashed one to 95\%. }%
\label{fig:YS}%
\end{figure}

Expansion of the pion wave function in terms of the Gegenbauer
polynomials has been defined in Eq.(\ref{gegen}) and the values of the
expansion coefficients for $\phi_{\pi}(u)$ of Eq.(\ref{Phiu}) are displayed in
Table \ref{tb:gegen}. Schmedding and
\begin{table}[h]
\caption{
The values of $<k_{T}^{2}>$, $<k_{T}^{4}>$ and
$\frac{<k_{T}^{4}>}{<k_{T}^{2}>^{2}}$ }%
\label{tb:ktav}
\begin{center}
\begin{tabular}[c]{ccccc}%
$M$ & n & $<k_{T}^{2}>$ & $<k_{T}^{4}>$ & ${< k_{T}^{4} >}/ {< k_{T}^{2}>^{2}}$\\
MeV &  & GeV$^{2}$ & GeV$^{4}$ & \\\hline
& 1 & $(0.451)^{2}=0.203$ & $(0.620)^{4}=0.148$ & 3.58\\
325 & 2 & $(0.433)^{2}=0.188$ & $(0.564)^{4}=0.101$ & 2.90\\
& 5 & $(0.424)^{2}=0.180$ & $(0.542)^{4}=0.086$ & 2.67\\\hline
& 1 & $(0.437)^{2}=0.191$ & $(0.593)^{4}=0.124$ & 3.39\\
350 & 2 & $(0.420)^{2}=0.176$ & $(0.541)^{4}=0.086$ & 2.76\\
& 5 & $(0.411)^{2}=0.169$ & $(0.520)^{4}=0.073$ & 2.55\\\hline
& 1 & $(0.426)^{2}=0.182$ & $(0.572)^{4}=0.107$ & 3.24\\
375 & 2 & $(0.409)^{2}=0.167$ & $(0.522)^{4}=0.074$ & 2.65\\
& 5 & $(0.402)^{2}=0.162$ & $(0.503)^{4}=0.064$ & 2.45\\\hline
& 1 & $(0.418)^{2}=0.175$ & $(0.555)^{4}=0.095$ & 3.11\\
400 & 2 & $(0.402)^{2}=0.162$ & $(0.508)^{4}=0.067$ & 2.55\\
& 5 & $(0.394)^{2}=0.155$ & $(0.489)^{4}=0.057$ & 2.37
\end{tabular}
\end{center}
\end{table}
Yakovlev \cite{YakSch} performed recently
a theoretical analysis of the latest CLEO \cite{Cleo} data on pion
electroproduction from a quasi real photon. They constructed a 95\% and 68\%
confidence level contour plots in the $a_{2}-a_{4}$ parameter space for an
average virtuality $\mu=2.4$ GeV. In Fig.\ref{fig:YS} we present their contours
(Fig.6 in \cite{YakSch}) together with our values of $a_{2}$ and $a_{4}$ for
various fits displayed in Table \ref{tb:gegen}. We have not performed the QCD
evolution of these coefficients since it is not clear which scale should be
taken as a starting point $\mu_{0}$ of the evolution. It is naturally to
assume that $\mu_{0}\sim\Lambda$. In that case, however, the evolution would
be very weak, since -- depending on the cut-off function -- $\Lambda$ changes
between 1 and 3 GeV. As the result of the evolution the points in the
$a_{2}-a_{4}$ plane would slightly move towards the asymptotic wave function.

Let us finally note, that the analysis of Ref.\cite{YakSch} has been performed
under the assumption that $a_{6}=0$. As can be seen from Table \ref{tb:gegen}
for $n>1$ $a_{6}$ is still comparable with $a_{4}$.

\section{Summary and conclusions} \label{sumc}

In this paper we have calculated both $u$ and $k_{T}^{2}$
dependence of the light cone twist 2 pion wave function in the
non-local NJL model. The nonperturbative nature of this approach
is embodied in the momentum dependence of the constituent quark
mass. In principle this dependence is known from the instanton
model of the QCD vacuum. However, in order to be able to carry out
the calculations in the Minkowski space-time we have used a class
of simple Ans\"{a}tze (\ref{Fkdef}), which for Euclidean momenta
relatively well reproduce $M_{inst.}(k)$. In this way we have
constructed a transparent and tractable model which allowed to
calculate the pion wave function $\phi_{\pi}(u)$ analytically. We
have shown how for the finite cut-off parameter $\Lambda$ the
Feynman integration contour has to be modified in order that
$\phi_{\pi}(u)$ is real and vanishes outside $0<u<1$.

Our phenomenological results are encouraging. For the constituent masses at
zero momentum between $M=325-350$ MeV and $n=2$ (see Eq.(\ref{Fkdef})) our
results nicely fit into the 95\% confidence region of Ref.\cite{YakSch}.
In comparison with the asymptotic wave function (\ref{as}) and with the
results
of Refs.\cite{PetPob,Bochum} where the momentum
dependence of the constituent mass in the denominator (c.f. Eq.(\ref{Fpi1}))
was neglected, our wave function is broader with, however, less pronounced
end point regions. Similar behavior was found in Ref.\cite{BaMiSte} where not
only non-locality (within the sum rules approach) but also radiative
corrections have been taken into account. The fact that the true pion
distribution amplitude may be broader than the asymptotic one has been already
pointed out in Ref.\cite{NiSKim}.

Although at first sight we have found rather strong sensitivity to
the power $n$ in the $k_{T}^{2}$ distribution, it turned out that
the power like asymptotic tail
$\tilde{\phi}_{\pi}(k_{T}^{2})\sim(1/k_{T}^{2})^{4n+1}$ switches
on at rather high momenta of the order of a few GeV. For small and
moderate $k_{T}^{2}$ there is no dependence of
$\tilde{\phi}_{\pi}(k_{T}^{2})$
on $n$. We have shown that more than 98\% of the norm of $\tilde{\phi}_{\pi}(k_{T}%
^{2})$ is concentrated in the region of $k_{T}^{2}<1$ GeV$^{2}$. Therefore
recent measurements of the diffractive dissociation of a pion into 2 jets with
transverse momentum above $1$ GeV \cite{E791} test the tiny portion of the
wave function. Most probably, since the higher $k_{T}^{2}$ moments are
divergent for the power like tail, this region is not reliably described by
our model.

Another way to extract a light cone wave function would be to calculate
the entire physical process within the effective model, rather than only
the wave function itself as given by Eq.(\ref{Fipidef}). This approach has
been advocated recently in Ref.\cite{ADT} and for the process
$\gamma \gamma \rightarrow \pi$ can be in principle carried
out also within our approach. It would be interesting to see what is the
difference between the wave function calculated directly from
Eq.(\ref{Fipidef}) and according to the prescription of Ref.\cite{ADT}.

As a by-product we have also calculated the transverse photon wave function
$\phi_{\gamma}^{\bot}(u)$ which is almost insensitive to the parameter $n$ of
the constituent quark mass. Further application of our method to strange
mesons and, for example, 2$\pi$ distribution amplitudes is straightforward and
will be discussed elsewhere. It is also possible by calculating pion wave
functions with other Dirac structures in (\ref{Fipidef}) to disentangle
different condensates related to the $k_{T}^{2}$ moments \cite{Zhitkt}.

\vspace{0.3cm}

This work was partially supported by the Polish KBN Grant PB~2~P03B~{\-}%
019~17. M.P. is grateful to W.Broniowski, K.Goeke, P.V.Pobylitsa, M.V.Polyakov
and N.G.Stefanis for discussions and interesting suggestions. The authors
kindly acknowledge comments and remarks of A.Bakulev, A.Dorokhov and T.Heinzl.


\end{document}